\newcommand{\be}{\begin{eqnarray}}
\newcommand{\ee}{\end{eqnarray}}
\newcommand{\nn}{\nonumber \\}
\newcommand{\V}{{{\cal V}}}
\newcommand{\Q}{{{\cal Q}}}

\newcommand{\openone}{\mbox{1\kern -0.25em I}}
\newcommand{\openK}{\mbox{I\kern -0.25em K}}
\newcommand{\openZ}{\mbox{Z\kern -0.4em Z}}
\newcommand{\openR}{\mbox{I\kern -0.25em R}}
\newcommand{\openH}{\mbox{I\kern -0.25em H}}
\newcommand{\openM}{\mbox{I\kern -0.25em M}}
\newcommand{\openC}{\mbox{C\kern -0.55em I\hspace{0.25em}}}
\newcommand{\Id}{\mbox{I\kern -0.1em d}}

\newcommand{\til}{\tilde{~}}
\documentstyle[a4]{article}

\begin{document}
\title{Dirac theory from a field theoretic point of view}
\author{Bertfried Fauser\\
Institut f\"ur Theoretische Physik\\
Auf der Morgenstelle 14\\
72072 T\"ubingen {\bf Germany}\\
Electronic mail: Bertfried.Fauser@uni-tuebingen.de
}
\date{May 26, 1996}
\maketitle
\begin{abstract}
Several complications arise in quantum field theory because of
the infinite many degrees of freedom. However, the
distinction between one-particle and many-particle effects  -- 
mainly induced by the vacuum -- is not clear up to now. A field
theoretic picture of the one-particle Dirac theory is developed
in order to explore such questions. Main emphasis is laid on the
injection of Grassmann's algebra into the endomorphism Clifford
algebra built over it. The obtained ``field theoretic''
functional equation behaves in a very unusual way. New methods
to handle Dirac and QFT are given.
\end{abstract}

\section{Introduction}

The development of atomic physics was done in several steps,
which were adopted to fit the measurements of the contemporary
experiments. Thus from Bohr, Bohr--Sommerfeld to Pauli and Dirac
\cite{Bohr,Bohr-Sommerfeld,Pauli,Dirac}
there is an increasing accuracy in analogy to the development of
a high precise art doing spectroscopic experiments.
Beside several improvements concerning additional interactions,
there was a need of new concepts also. The discovery of spin and
anti-particles was not only a minor change in theory, it was the
birth of new concepts in atomic physics.
The fact, that the Dirac theory was able to calculate the
spectrum of the Hydrogen atom very well can not hide two major
drawbacks further present in this theory.

First, the Dirac theory accounts only for the anomalous
g--factor 2 and not for the small deviation $(g-2)=\alpha/\pi
+{\cal O}(\alpha^2)$, which therefore was calculated in quantum
electrodynamics (QED), by means of perturbation theory.

Second, despite of the outstanding triumph to predict correctly
the appearance of antiparticles, which were not even known in
1928, the mere fact of their introduction spoils the theory to
be a one-particle theory by Dirac-sea issues.

Furthermore, one obtains by introducing a QFT vacuum, which
exhibits stochastic fluctuations, the Casimir effect. One
calculates therewith a force in the empty space of a cavity due
to ``vacuum polarisation''.

The problems seem to be cured in the most accurate present day
theory of physics, namely QED. But the necessary(?) field
quantization of the gauge field and the so called ``second
quantization'' of the fermion field are another revolutionary
development of the theory. There is no commonly accepted way
back from QED to Dirac theory.

Now, present day developments in Clifford algebra allow a much
more profound reexamination of that important connection. The
new calculations, done by Kr\"uger \cite{Krue}, with surprising
results, show, that something may have been overlooked in the
development of Dirac's theory. An exiting evolutionary
generalization was promoted by Daviau \cite{Daviau}. Originated
in a thoughtful study of the algebraic properties of Dirac's
equation, he obtained a non-linear Dirac equation, a special
case of Lochak's monopole equation \cite{Lochak}, which is called
Dirac--Daviau equation. This equation exhibits several new and
useful properties \cite{Daviau-thesis}. Daviau was able to show,
that his non-linear equation provides for the Hydrogen atom a
nearly analogous spectrum as the linear one, with the additional
feature, that the small correction may account for the Lamb
separation of otherwise degenerate energy levels. 

Now, to turn the way around, one may argue, that it should be
possible to find a way back from QED to Dirac's or Daviau's
theory. Since it is possible to formulate QED in terms of
Clifford algebras \cite{Fauser-pos}, this opens a new
possibility to connect the two adjacent theories. Moreover, we
hope to understand in this way how to manage the distinction of
one-particle and multi-particle effects. This is nothing else,
as to find an alternative explanation of the so called ``vacuum
fluctuations''. As a first step, we try to find a formulation of
the Dirac equation, which exhibits as much as possible the field
theoretic features {\it without}\/ leaving the one-particle picture.

\section{Clifford algebra as endomorphism algebra of a certain
Grassmann algebra}

{\bf Definition:} The Clifford algebra $CL(\V,\Q)$ of the pair
$(\V,\Q)$ of a linear (vector) space $\V$ and a quadratic form
$\Q$ is the universal algebra obtained by the linear Clifford
map $\gamma$ satisfying
\be\label{clifford-map}
\gamma : \V & \longrightarrow & CL(\V,\Q) \nn
\V \ni x & \longrightarrow & \gamma_x\nn
&& \gamma_x \gamma_x = \gamma_{x}^2 =\Q(x) \nn
&& CL(\V,\Q) = <\Id , \gamma_{x_i},
 \gamma_{x_i}\gamma_{x_j,i<j}, \ldots> 
 \mbox{~~with~~} \V=<x_i>. \Box
\ee
One might say, that a Clifford algebra is the algebra compatible
to a quadratic form. Thereby, every Clifford algebra is related
to the geometry induced by the quadratic form $\Q$. In the sequel,
we will deal exclusively with non-degenerate quadratic forms.
The linear (vector) space $\V$ is assumed to be build over the
fields \openR\ or \openC, denoted by \openK.

The usual way to introduce the Clifford algebra is to polarize
the above Clifford map $\gamma : \V \rightarrow CL$, to obtain
the usual commutation relations. Therefore one has the following\\
{\bf Definition:} The polar form $B_p$ is the symmetric bilinear
form related to $\Q$ by
\be
B_p : \V \times \V & \longrightarrow & K \\
2B_P(x,y) & = & \Q(x+y)-\Q(x)-\Q(y). \hfill\Box
\ee
The factor 2 was introduced for convenience and causes no
problems, because our field \openK\ is of $char \not= 2$~$(char
\openK =0)$. 

With a $\Q$ $(B_p)$ orthonormal set of generators $\{ e_i\}$,
one obtains from (\ref{clifford-map}) the usual commutator
relations 
\be
\gamma_{e_i}\gamma_{e_j}+\gamma_{e_j}\gamma_{e_i}
&=& 2 B_p(e_i,e_j) = 2 diag(+1,\ldots,-1).
\ee

A widely used identification is, to drop the distinction between
the linear (vector) space $\V$ and its image $\V(p,q)$ or $\V(n)$
inside the Clifford algebra. We use the term vector space to
denote $\V(p,q)$ in $CL$ and linear space for $\V$. Note, that one
can not speak of a signature without looking at the pair $(\V,\Q)$
over \openR, which is by the universality property essential
equivalent to speak about the Clifford algebra
$CL(\V,\Q)\equiv CL(p,q)$. The same identification is done with the
field \openK\ itself, identifying \openK\ with $\gamma_{\openK}
\subset CL(\V,\Q)$.

A Clifford algebra possesses two important involutions, which are
usually given on the set of orthonormalized generators and
afterwards on the whole algebra by linearity.\\
{\bf Definition:}
The main-involution is given by
\be
\hat{~} &:& CL \longrightarrow CL \nn
&& \gamma_{e_i} \rightarrow \hat{\gamma}_{e_i}=-\gamma_{e_i},
\ee
while the main-anti-automorphism (involution), also called
reversion, is given by
\be
\til &:& CL \longrightarrow CL \nn
&& \gamma_{e_i} \rightarrow \gamma\til_{e_i}=\gamma_{e_i} \nn
&& \gamma_{e_i}\gamma_{e_j} \rightarrow 
  (\gamma_{e_i}\gamma_{e_j})\til =
  \gamma_{e_j}\til\gamma_{e_i}\til =
  \gamma_{e_j}\gamma_{e_i},\quad \mbox{etc.}
\ee
We define the conjugation $\bar{~}$ as $\bar{~}:=\til\circ
\hat{~} = \hat{~}\circ\til.$ \hfill $\Box$.

It will become clear later, that a wedge product often
introduced by many authors depends on the
bilinear form of a Clifford algebra including a possible
non-symmetric part. The $\openZ_2$--grading of a Clifford
algebra is obtained from the main-involution, $CL=CL^+\oplus
CL^-$, and thus is not $B_p$ or $B$ dependent.

Now, in employing this identification, we have done much more.
Knowing a bilinear form $B_p$ on $\V$, is equivalent to know an
adjoint map from $\V$ to $\V^\ast$, the space of linear forms of
$\V$, which itself is a linear (vector) space.

{\bf Definition:} The adjoint map $\ast_{B_p}$ related to the
polar bilinearform $B_p$ of $\Q$ is given by
\be
\ast_B : \V & \longrightarrow & \V^\ast \nn
&& \V^\ast(p,q)\ni x^{\ast_B} := B(x,\cdot) \nn
&& x^{\ast_B}(y) = B(x,y). \Box
\ee
Now, it is well known, that there is no {\it canonical}\/ dual
isomorphism, which relates $\V$ and $\V^\ast$ as linear spaces.
But, as we have seen, the Clifford algebra provides us with such
a dual isomorphism by its construction. Thus we are dealing
with the pair ($CL(\V,\Q), \ast_B)$, the Clifford algebra and
the dual isomorphism.

As we are studying the finite dimensional case, no difficulties
should occur, but it is exactly this step, which is crucial in
field theory. The dual space $\V^\ast$ of an infinite dimensional
space $\V$ may be of larger cardinality as $\V$. In this case, we
are not able to identify $\V$ with $\V^{\ast\ast}$ in a canonical
way.

But have we done the thing right already in the finite theory?

Because the Clifford algebra is the universal algebra of a
linear (vector) space and a quadratic form, we have to look, if
every non-singular bilinear form, which gives us a dual
isomorphism is obtained in this way. The answer is the following\\
{\bf Proposition:} The symmetric part of a non-singular bilinear
form determines the corresponding quadratic form uniquely. \hfill$\Box$

Thus the Projection from the space of bilinear forms onto the
space of quadratic forms is not surjectiv,
\be
\frac{\mbox{bilinear forms}}{\mbox{alternating forms}}
& \cong & \mbox{quadratic forms}.
\ee
Now, not being able to construct a unique bilinear form from a
quadratic form, we are still {\it not}\/ able to construct a dual
isomorphism. 

Since quantum mechanics, as quantum field theory in a much more
elaborated way, is based on states and duality in the sense of
an adjoint, we are not jet able to study quantum mechanics with
Clifford algebras. But on the other hand, simple one-particle
quantum mechanics does fix the choice immediately and so also
omits this freedom.

In the Hestenes formulation of Dirac's theory \cite{Hestenes} we
have simply
\be
\vert \psi> & {DH \atop ~~\longrightarrow~~} & \Psi \in CL^+ \nn
<\psi \vert & {DH \atop ~~\longrightarrow~~} & \tilde\Psi \in CL^+.
\ee
Thus the Dirac adjoint is represented by the reversion, which
was defined on the orthonormalized generators of $CL(\V,\Q)$.

To be able to give a more elaborate answer to the above
question, we will construct the Clifford algebra from the
Grassmann algebra by Chevalley deformation. This process
directly relates an arbitrary bilinear form $B$ to the pair
$(CL(\V,\Q),\ast_B)$ in a unique way.

{\bf Definition:} The Grassmann algebra $\Lambda[\V]$ of the
linear (vector) space $\V$ is the universal algebra obtained by
the linear injection $j$ with the properties
\be
j : \V & \longrightarrow & \Lambda[\V] \nn
\V \ni x & \longrightarrow & j_x \nn
&& j_x \wedge j_x = j_{x}^2 =0.
\ee
The product in $\Lambda[\V]$ is often denoted by the $\wedge$
(wedge product). \hfill$\Box$

Remark: Because of the fact, that the wedge product, called
``combinatorial product (Kombinatorisches Produkt)'' by
Grassmann \cite{A2}, describes a disjoint {\it union}\/
one should use the vee product $\vee$ instead. 

The Grassmann algebra of a finite dimensional linear (vector)
space $\V$, $dim\V=n$, is a $\openZ_n$--graded finite (unital,
associative \openK--) algebra of dimension $2^n$. Each element
of $\Lambda[\V]$ can be decomposed into parts of homogeneous
degree. Thus 
\be
\Lambda[\V] &=& \oplus_{r=0}^n \Lambda^r[\V] \nn
dim: 2^n    &=& \sum_{r=0}^n \left( {n \atop r } \right).
\ee
Given a set $\{ x_i \}$, which generates $\V$ as linear space,
every element $X$ in $\Lambda[\V]$ can be written as 
(summation convention employed)
\be
X &=& \alpha_0 \Id+\alpha_i j_{x_i} + \alpha_{ij} j_{x_i}\wedge
j_{x_j,i<j}+ \ldots \nn
&& \{\alpha_0,\alpha_i,\ldots\} \in \openK.
\ee
Obviously, $\Lambda[\V]$ can also be decomposed into even and
odd parts by the main-involution
\be
\hat{~} &:& \V \longrightarrow -\V \nn
&& \hat{j}_{x_i}=-j_{x_i} \quad \forall j_{x_i}\quad \in \V \nn
&& \Lambda[\V]=\Lambda^+[\V] \oplus \Lambda^-[\V].
\ee
To reobtain the Clifford algebra, we introduce the dual space of
linear forms $\V^\ast$ of the linear (vector) space $\V$. With
the same generating set $\{ x_i \}$ of $\V$ we have the\\
{\bf Definition:} The dual base in $\V^\ast$ is given by the
co-vectors $\partial^{x_i}$ subjected to the conditions
\be
\partial^{x_i} j_{x_j} &=& \delta^i_j. \hfill\Box
\ee
In this way the action of every co-vector element $a\in \V^\ast$
on the vector elements $x\in\V$ is given. But to ensure the
action of the co-vectors on multi-vectors out of $\Lambda[\V]$,
we have to require, that the $\partial^x$ are {\it
anti-derivations} of degree $-1$. In addition, we have to define
how to apply repeated derivations and the action of
multi-co-vectors. This leads to the well known properties of
$\partial^x$ on the generators, which by linearity are defined
on the whole algebra \cite{Oziewicz86} 
\be
\partial^{\alpha x_i+\beta x_j} = \alpha\partial^{x_i}
+\beta \partial^{x_j} \mbox{\hspace{1.5cm}}
&~& \mbox{linearity} \nn
\partial^{x_i}(j_{x_j}) = \delta^i_j \mbox{\hspace{1.5cm}}
&~& \mbox{def. of
co-vectors, $x_i\in \V$} \nn
\partial^{x_i} j_{x_{j_1}}\wedge \ldots \wedge j_{x_{j_n}} 
= \mbox{\hspace{1.5cm}}
&~& \mbox{graded Leibniz rule} \nn
\sum_{r=1}^n (-)^{r+1}\delta^i_{j_r}
j_{x_{j_1}}\wedge \ldots \wedge j_{x_{j_{r-1}}}\wedge
j_{x_{j_{r+1}}}\wedge \ldots \wedge j_{x_{j_n}} 
&~& \nn
\partial^{x_{i_1}\wedge \ldots \wedge x_{i_r}}
j_{x_{j_1}}\wedge \ldots \wedge j_{x_{j_n}} = \mbox{\hspace{1.5cm}}
&~& \mbox{left module structure} \\
\partial^{x_{i_1}}(\ldots (\partial^{x_{i_r}}(
j_{x_{j_1}}\wedge \ldots \wedge j_{x_{j_n}}))\ldots ) 
&~& (=det~\partial^{x_i} j_{x_j}\mbox{~if $\#x_i=\#x_j$})\nonumber
\ee
Therefore, we are able to build up from the dual space $\V^\ast$
also a Grassmann algebra, acting on $\Lambda[\V]$ in a
multilinear way. But one has to care not to confuse
$\Lambda[\V^\ast]=\Lambda[<\partial^{x_i}>]$ with $[\Lambda[\V]
]^\ast$. 

If we look at the endomorphisms of $\Lambda[\V]$, we have the
following\\
{\bf Theorem:} (Greub \cite{Greub,Rodrigues})
\be
End(\Lambda[\V]) 
&\cong& 
CL(\V\oplus\V^\ast,\delta\perp \delta) 
\cong
CL(\V\oplus\V ,\delta\perp -\delta) 
\cong
\Lambda[\V]\otimes\Lambda[\V^\ast] \nn
&&\delta(x_i) = x_i x_i =1,\quad \V=<x_i>,\quad
\V^\ast=<\partial^{x_i}>. \hfill\Box
\ee
In another language, we may look at the elements of the
Grassmann algebra as {\it states\/} and at the elements of the
Clifford algebra as {\it operators\/}, acting by left
multiplication on states.

To make this connection more explicit, we give the\\
{\bf Definition:}
\be\label{def-gamma}
\gamma^{x_i} &:=& \sum_j \delta^{x_ix_j}j_{x_j}+\partial^{x_i} \nn
\gamma^{\ast {x_i}} &:=& \sum_j
\delta^{x_ix_j}j_{x_j}-\partial^{x_i} 
\Box
\ee
{}From this we can easily calculate the commutation relations
\be
\gamma^{x_i}\gamma^{x_j}+
\gamma^{x_j}\gamma^{x_i} &=& ~~2\delta^{ij} \nn
\gamma^{\ast {x_i}} \gamma^{\ast {x_j}}+
\gamma^{\ast {x_j}} \gamma^{\ast {x_i}} &=& -2\delta^{ij} \nn
\gamma^{x_i}\gamma^{\ast {x_j}}+
\gamma^{\ast {x_j}} \gamma^{x_i} &=& ~~~0.
\ee
Of course, this construction is a cover of $End(\Lambda[\V])$,
which is generated by the $\gamma^{x_i}$ and $\gamma^{\ast x_i}$.
Nevertheless, by dimensional arguments, we have
$dim~\Lambda[\V]=2^n$ and $dim~End[\Lambda[\V]] = 2^{2^n}$
and need therefore both kinds of $\gamma$'s to
establish the connection between the $\gamma$ and the
$j,\partial$ picture. We may invert (\ref{def-gamma}), which is
possible for every non-degenerate symmetric bilinear form
(non-degenerate $\Q$) and obtain (s.c. employed)
\be
j_{x_i} &=&
\frac{1}{2}\delta_{x_ix_j}(\gamma^{x_j}+\gamma^{\ast {x_j}}) \nn 
\partial^{x_i} &=& \frac{1}{2}(\gamma^{x_i}-\gamma^{\ast {x_i}}).
\ee
If we introduce the involution $\ast$ by 
\be
\ast &:& CL(\V\oplus \V^\ast,\delta\perp \delta) 
\longrightarrow 
CL(\V\oplus \V^\ast,\delta\perp \delta) \nn
&& \gamma~~\rightarrow~~\gamma^\ast \nn
&& \gamma^\ast~\rightarrow~~\gamma,
\ee
we obtain $\Lambda[\V]$ as the space, which is stabilized by
this $\ast$--involution.

Because of the \\
{\bf Theorem:} (Chevalley, \cite{Chevalley})
\be
CL(\V,-\Q) &\cong & CL(\V,\Q)^{op}, \Box
\ee
where $CL^{op}$ is the algebra with reversed products, we can
identify the $\gamma^{\ast {x_i}}$ as a $\gamma^{x_i}$ acting from
the right. Set $u= \gamma^{x_1}\ldots\gamma^{x_n}$, then we have
(see also \cite{AblamowiczLounesto})
\be
\ast &:& Cl \longrightarrow Cl^{op} \nn
&& \ast(\gamma^{x_k}u) = \hat{u}\gamma^{\ast {x_k}}
\ee
Thus, due to the necessity of working with (vector) states in
quantum mechanics, we have to take care to simulate the right
action of $\gamma^{x_i}$ by the left action of
$\gamma^{\ast {x_i}}$. The involution $\ast$ may be looked at as a
sort of transposition or (Hermite) adjoint.

Now, the same construction is possible for an arbitrary
non-degenerate not necessary symmetric bilinear form $B$, with
symmetric part $A$ and anti-symmetric part $F$. The matrices of
these bilinear forms on the generating set $\{x_i\}$ will be
denoted by $[B^{ij}]=[B(x_i,x_j)]$. We end up with the following
formulas (s.c. employed)
\be\label{def-neu-gamma}
B^{op}(x,y) &=& -B(y,x) = -A+F \nn
\gamma^{x_i} &:=&[B^{ij}]j_{x_j}+\partial^{x_i}=
[A^{ij}]j_{x_j}+[F^{ij}]j_{x_j}+\partial^{x_i} \nn
\gamma^{\ast {x_i}} &:=&[B^{op~{ij}}]j_{x_j}-\partial^{x_i}=
-[A^{i,j}]j_{x_j}+[F^{ij}]j_{x_j}-\partial^{x_i} \nn
{}
[A^{ij}] &\equiv & [A_{ij}^{-1}] \quad\mbox{(exists by assumption)} \nn
j_{x_i} &:=&
\frac{1}{2}[A_{ij}](\gamma^{x_j}+\gamma^{\ast {x_j}}) \nn
\partial^{x_i} &:=&
\frac{1}{2}(\gamma^{x_i}-\gamma^{\ast {x_i}})-
\frac{1}{2}[F^{ij}][A_{jk}](\gamma^{x_k}+\gamma^{\ast {x_k}}).
\ee
We drop in further formulas the brackets $[\cdot ]$. It is
obvious, that this sort of {\it new}\/ $\gamma$'s are subjected to
new commutation relations 
\be
\gamma^{x_i}\gamma^{x_j}+
\gamma^{x_j}\gamma^{x_i} &=& ~~2A^{ij} \nn
\gamma^{\ast {x_i}}\gamma^{\ast {x_j}}+
\gamma^{\ast {x_j}}\gamma^{\ast {x_i}} &=& -2A^{ij} \nn
{\gamma}^{x_i}\gamma^{\ast {x_j}}+
\gamma^{\ast {x_j}}{\gamma}^{x_i} &=& ~~~0,
\ee
and that the involution $\ast$ does no longer stabilize the
space spanned by the $j$--sources. Hence, this transformation
can no longer be identified by a matrix transposition.
Furthermore, the alternating part $F$ is completely absent in
this commutators. 

We have to find a meaning of the norms, which are important
to describe ``scalar products''. The usual way
\cite{Porteous,Maks} is, to study the fields (double fields)
real Clifford algebras are built over. With the notations
${}^2\openR=\openR \oplus\openR$ and $\openR(2)=Mat_{2\times
2}(\openR)$, we have the relevant left modules built over
\be
M &:=& \{ \openR,\openC,\openH,{}^2\openR,{}^2\openH \}.
\ee
These modules are the spaces of (double) spinors, only one of
them equals $\openC$. 

We may utilize two different anti-automorphisms to build up
scalar products.\\
{\bf Definition:}
\be\label{def-alpha}
\alpha_\varepsilon 
&:=& 
\left\{
\begin{array}{ccl}
\til & \Leftrightarrow & \varepsilon = +1 \\
\bar{~}=\til\circ \hat{~} & \Leftrightarrow & \varepsilon = -1 \Box
\end{array} 
\right. 
\ee
We have thus $\alpha_\varepsilon(\V(p,q)) = \varepsilon\V(p,q)$
etc. With help of this map we may build a form
$\Phi_\varepsilon$ by
\be
\Phi_\varepsilon &:& CL(\V,\Q) \times CL(\V,\Q)
\longrightarrow CL(\V,\Q) \nn
&& \Phi_\varepsilon(r_1,r_2)=\alpha_\varepsilon(r_1)r_2.
\ee
For any $g\in CL(\V,\Q)$, we have the left $CL(\V,\Q)$--modul
homomorphism 
\be
g &:& CL(\V,\Q) \longrightarrow CL(\V,\Q) \nn
&& g(r)=gr,
\ee
and the {\it adjoint}\/ w.r.t. $\Phi_\varepsilon$ is given by
$\alpha_\varepsilon (g)$. In this way, the form
$\Phi_\varepsilon$ induces the map
\be
\mbox{N}_\varepsilon &:& CL(\V,\Q) \longrightarrow CL(\V,\Q) \nn
&& \mbox{N}_\varepsilon (r) = \alpha_\varepsilon(r)r
=\Phi_\varepsilon (r,r) \nn
&& \mbox{N}_\varepsilon(v)=\varepsilon v^2=\varepsilon\Q(v),\quad
v\in\V(p,q).
\ee
The invariance group of $\Phi_\varepsilon$ (under
multiplication) is given by
\be
Inv_\varepsilon(p,q) &=&
\{ g\in CL(\V,\Q) \vert \mbox{N}_\varepsilon(g)=1 \}.
\ee
Lounesto has shown, that there exists always a form equivalent
(isomorphic) to $\Phi_\varepsilon$, such that the restriction of
$\mbox{N}_\varepsilon$ to (double)-- spinors is $M$--valued 
\cite{Lounesto5}. The group of automorphisms, which preserve the
``scalar product'' $\Phi_\varepsilon$ is thus isomorphic to
$Inv_\varepsilon(p,q)$.\\
{\bf Definition:} (Lipschitz group) Let
$CL^\times(\V,\Q)$ denote the units in $CL(\V,\Q)$ w.r.t.
$\Phi_\varepsilon$. The Lipschitz group $\Gamma(p,q)$ is given
as 
\be
\Gamma(p,q) &:=& 
\{r\in CL^\times(\V,\Q) \vert r \V(p,q) \hat{r}^{-1} \subset
\V(p,q) \}. \Box
\ee
For every $r\in \Gamma(p,q)$,
\be
\rho_r &:& \V(p,q) \longrightarrow \V(p,q) \nn
&& \rho_r(v) = rv\hat{r}^{-1}
\ee
belongs to ${\bf O}(p,q)$. Further we have the important\\
{\bf Proposition:}
The pseudo--norm $\mbox{N}_\varepsilon$ is $\openK$ valued on
$\Gamma(p,q)$. \hfill $\Box$

Because one can write every element $r$ of $\Gamma(p,q)$ as
$r=v_1\ldots v_n$, with $v_i\in \V(p,q)$, we obtain
\be
\mbox{N}_\varepsilon &=&\alpha_\varepsilon(r)r=
\varepsilon^n\Q(v_1)\ldots \Q(v_n),
\ee
which is in $\openK$.

Thus, our above construction involves an anti-automorphism and
thereby induce in general a $M$--valued (degenerate) form on the
whole algebra. We have thus to search for ``states'', which are
in $\Gamma(p,q)$ (pure states), to obtain $\openK$--valued
``expectation values''. 

Before we proceed to study Dirac's theory with $j$--sources and
Chevalley deformation, we consider the conjugations. From the
linear dependence of $\gamma,\gamma^\ast$ from $j,\partial$
(\ref{def-neu-gamma}) we have  
\be
\hat{\gamma}^{x_i} &=& -(\gamma^{x_i})
=-(\partial^{x_i}+B^{ij}j_{x_j}) \nn
&=&(-\partial)^{x_i}+B^{ij}(-j)_{x_j} 
\ee
and thus
\be
\hat{\partial}^{x_i} &=& -\partial^{x_i}\nn
\hat{j}_{x_i} &=& -j_{x_i}.
\ee
{}From 
\be
\gamma^{x_k}\gamma^{x_i}
&=&
\partial^{x_k}\partial^{x_i}+B^{ik}-B^{il}j_{x_l}\partial^{x_k}
+B^{kj}j_{x_j}\partial^{x_i}+B^{ij}B^{kl}j_{x_l}j_{x_j}\nn
&=&
(\partial^{x_i}\partial^{x_k}+B^{ik}-B^{il}\partial^{x_k}j_{x_l}
+B^{kj}\partial^{x_i}j_{x_j}+B^{ij}B^{kl}j_{x_j}j_{x_l})\til\nn
&=&
(\partial^{x_i}\partial^{x_k}+B^{ik}-B^{ik}+B^{kl}j_{x_l}\partial^{x_i}
+B^{ki}-B^{il}j_{x_l}\partial^{x_k}+B^{ij}B^{kl}j_{x_j}j_{x_l})\til\nn
&=&
(\gamma^{x_i}\gamma^{x_k})\til,
\ee
we conclude that $\til$ reverse products of $j$ and $\partial$. Thus
\be
(j_{x_{r_1}}\ldots j_{x_{r_n}}
 \partial^{x_{s_1}}\ldots\partial^{x_{s_m}})\til &=&
 \partial^{x_{s_m}}\ldots\partial^{x_{s_1}}
 j_{x_{r_n}}\ldots j_{x_{r_1}}.
\ee
e.g.
\be
(j_{x_i}\partial^{x_k})\til &=& \partial^{x_k}j_{x_i}
 =\delta^k_i-j_{x_i}\partial^{x_k}\nn
 (j_{x_i}\partial^{x_j}\partial^{x_k})\til &=&
 \delta^j_i\partial^{x_k}-\delta^k_i\partial^{x_j}
 +j_{x_i}\partial^{x_k}\partial^{x_j}.
\ee
The star conjugation results in
\be
j^\ast_{x_i} &=&
[\frac{1}{2}A_{ij}(\gamma^{x_j}+\gamma^{\ast {x_j}})]^\ast \nn
&=&\frac{1}{2}A_{ij}(\gamma^{\ast {x_j}}+\gamma^{x_j})
=j_{x_i} \nn
\partial^{\ast {x_i}} &=&
[\frac{1}{2}(\gamma^{x_i}-\gamma^{\ast {x_i}})
-\frac{1}{2}F^{ij}A_{jk}(\gamma^{x_k}+\gamma^{\ast {x_k}})]^\ast \nn
&=&-\frac{1}{2}(\gamma^{x_i}-\gamma^{\ast {x_i}})
+\frac{1}{2}F^{ij}A_{jk}(\gamma^{x_k}+\gamma^{\ast {x_k}})
-F^{ij}A_{jk}(\gamma^{x_k}+\gamma^{\ast {x_k}}) \nn
&=&
-\partial^{x_i}-2F^{ij}j_{x_j} \nn
\partial^{\ast\ast {x_i}} &=& -\partial^{\ast {x_i}}-2F^{ij}j_{x_j}
=\partial^{\ast {x_i}}+2F^{ij}j_{x_j}-2F^{ij}j_{x_j} \nn
&=& \partial^{x_i}
\ee
This suggests to introduce the new derivative
\be
\mbox{d}^{x_i} &:=&\partial^{x_i}+F^{ij}j_{x_j} \nn
\mbox{d}^{\ast {x_i}} &=& -\partial^{x_i}-2F^{ij}j_{x_j}
+F^{ij}j_{x_j} 
=-(\partial^{x_i}+F^{ij}j_{x_j})=-\mbox{d}^{x_i},
\ee
which is thus $\ast$ anti-stable. Of course, exact this was the
motivation to introduce normal ordering in QFT
\cite{Fauser-vertex}. 
The commutator relations are derived to be
\be
\{\mbox{d}^{x_i},\mbox{d}^{x_j} \} &=& ~0 \nn
\{\mbox{d}^{x_i},j_{x_j} \} &=& ~\delta^i_j \nn
\{ j_{x_i}, j_{x_j} \} &=& ~0.
\ee
Thus we obtain a new set of co-vectors, subjected to the {\it
same}\/ commutation relations. But, the star conjugation (anti)
stabilizes the $\mbox{d},j$ parameterization and {\it not}\/ the
$\partial,j$ set. One reobtains in the $\mbox{d},j$ picture the
usual matrix transposition. The substitution $\partial
\rightarrow \partial+Fj$ is exactly the functional form of
normal ordering in QFT, if one asserts $F$ to be the propagator
\cite{Fauser-thesis,Fauser-pos}. 

For defining the adjoint, we combine the reversion and
conjugation again into $\alpha_\varepsilon$ (\ref{def-alpha}).
We obtain the action of $\alpha_\varepsilon$ as
\be
\alpha_\varepsilon(j_{x_i}) &=& \varepsilon j_{x_i} \nn
\alpha_\varepsilon(\partial^{x_i}) &=& \varepsilon
\partial^{x_i} \nn
\alpha_\varepsilon(j_{x_1}\ldots j_{x_n}\partial^{x_{n+1}}
\ldots \partial^{x_{n+r}}) &=&
\alpha_\varepsilon(\partial^{x_{n+r}})\ldots
\alpha_\varepsilon(\partial^{x_{n+1}}) 
\alpha_\varepsilon(j_{x_n})\ldots \alpha_\varepsilon(j_{x_1}) \nn
&=&
\varepsilon^{n+r}\partial^{x_{n+r}}\ldots\partial^{x_{n+1}}
j_{x_n}\ldots j_{x_1}.
\ee
In low dimensional cases one obtains
\be
\alpha_\varepsilon(\Id ) &=& \Id \nn
\alpha_\varepsilon(j_{x_i}\partial^{x_k}) &=&
\varepsilon^2
\partial^{x_k}j_{x_i}=\delta^k_i-j_{x_i}\partial^{x_k} \nn
\alpha_\varepsilon(j_{x_i}j_{x_k}\partial^{x_l}) &=&
\varepsilon^3 \partial^{x_l}j_{x_k}j_{x_i} 
= \varepsilon(\delta^l_k j_{x_i}-\delta^l_i
j_{x_k}-j_{x_i}j_{x_k}\partial^{x_l}) \nn
&=& (\delta^l_k-j_{x_k}\partial^{x_l})\varepsilon(j_{x_i}) \nn
&=& \alpha_\varepsilon(j_{x_k}\partial^{x_l})
\alpha_\varepsilon(j_{x_i}).
\ee
The derivation structure is thus compatible with the action of
the adjoint. Hence, we are able to define a form on ``states'',
which are built of $j$--sources and a projector $\vert
0>_F<0\vert$ (see \ref{fock});
\be
\vert X> &=& \sum_{i,j,k,\ldots}(\alpha_0\Id + \alpha_i j_{x_i}
+ \ldots ) \vert 0>_F<0\vert\nn
<X\vert & := &\alpha_\varepsilon(\vert X >) \nn
\Phi_\varepsilon(X,Y) &:=& \alpha_\varepsilon(\vert X >) \vert
Y> = <X\vert Y> \in CL(p,q)
\ee
If we require the $\vert X>$ to be in the Lipschitz group
$\Gamma(p,q)$, we obtain a ``scalar product'' into the center of
$CL(p,q)$, which is $M$--valued.

Remark: Up to this state of the development, we are not able to
utilize the dual isomorphism $co_\V$,
\be
co_\V (j_{x_i}) &=& \partial^{x_i} \nn
co_\V (\partial^{x_i}) &=& j_{x_i},
\ee
because of the raising and lowering of the involved indices,
which could be done by $A$, $F$, $B$ or in another way. In a
Cartesian picture with $\partial^i=\partial_i$, $j^i=j_i$,
$co_\V$ would lead to a Fock space construction \cite{delanghe}.

\section{Dirac theory}

Dirac theory is usually given in terms of matrix
representations. For the purpose of reference, we use Bjorken
and Drell \cite{BjorkenDrell}, 
\be\label{Dirac}
\sum_{\beta=1}^4\left(
\sum_{\mu=0}^3(i\hbar \gamma^\mu_{\alpha\beta}
\frac{\partial}{\partial x^\mu}
+\frac{e}{c}\gamma^\mu_{\alpha\beta}A_\mu)
+mc\delta_{\alpha\beta}\right) \psi_\beta&=& 0,
\ee
where $\psi_\beta$ is a component of a column spinor, hence 
$\psi_\beta : \openM_{1,3} \longrightarrow \openC$. The
$\gamma$ matrices are elements of $Mat_{4\times 4}(\openC)$. The
spinor inner product is given by the Dirac adjoint
$\bar{\psi}^D_\beta =(\gamma^0\psi)^{\ast T}_\beta$, where
$\ast$ is complex conjugation and $T$ denotes transposition of
the matrix representation. Thus
\be
\sum_{\beta=1}^4\bar{\psi}^D_\beta\psi_\beta &=&\alpha \in \openR
\ee
is the standard scalar product. The polar bilinear form is just
the diagonal matrix $\eta^{\mu\nu}=diag(1,-1,-1,-1)$, which can
be seen from 
\be
\sum_{\beta=1}^4(\gamma^\mu_{\alpha\beta}\gamma^\nu_{\beta\rho}
+\gamma^\nu_{\alpha\beta}\gamma^\mu_{\beta\rho})
&=&
2\eta^{\mu\nu}\delta_{\alpha\rho}.
\ee
Now, there are several methods to establish a connection to
Clifford algebraic spinors. Since we are interested in
``states'', we prefer ideal spinors \cite{ideal-spinors},
because of their modul structure.

Let us start from an orthonormalized set of generating
elements $\{ e^\mu \}$, and the bilinear form
$B=B(e^\mu,e^\nu)=\eta^{\mu\nu}+F^{\mu\nu}$, with an arbitrary
function $F^{\mu\nu}$. This is no restriction since
$\eta^{\mu\nu}$ is constant, because we are able to
diagonalize the non-degenerate symmetric part of the bilinear
form, and afterwards normalize it to $\pm 1$ according to
Sylvester's theorem. This results thereby in a change of the
alternating part also!

We can built a primitive idempotent element $P$ as
\be
P & \equiv & P_{11}=
\frac{1}{2}(\Id+e^0) \frac{1}{2}(\Id+i e^2 e^3).
\ee
The appearance of a non-geometric $i$ is due to the fact, that
we are interested in a $Mat_{4\times 4}(\openC)$ representation
of the Dirac theory, which is quite artificial, since the Dirac
algebra $CL_{1,3}(\openR)\equiv CL(\openR^4,\eta)$ should be
represented in $Mat_{2\times 2}(\openH)$. Nevertheless, the
throughout appearance of $i\gamma^2$ has also physical
consequences. One might look at Dirac theory as a theory over
$CL_{2,2}(\openR^4)$, which is totally null and twistor like.
This algebra can be represented in $Mat_{4\times 4}(\openR)$. 
For example, angular momentum is studied most easily with
$l_\pm:=\gamma^1\pm i\gamma^2$, $l^3:=\gamma^3$ and $\gamma^5$
is usually defined as $\gamma^5:=i(\gamma^0\ldots\gamma^3)=
\gamma^0\gamma^1(i\gamma^2)\gamma^3$, which reflects our choice.

We define the set
\be
t_i&:=&\{ \Id,e^1e^3,e^3,e^1\},
\ee
from which we obtain
\be
\bar{t}_i=\alpha_{-1}(t_i)&=&\{ \Id,e^3e^1,-e^3,-e^1\}.
\ee
Since no wedge product is involved, this can be used in the case
$F^{\mu\nu}\not=0$ also. Then one has to care, not to involve
the common rules of reversion
\cite{Fauser-thesis,Fauser-pos,Fauser-mandel,AblamowiczLounesto}. 

Now we can build a base $P_{ij}$ of the matrix algebra, where
every $P_{ij}$ has only one $1$ at the $i$--th row and $j$--th
column and the other elements zero
\be
P_{ij}&:=& \bar{t}_i P t_j.
\ee
Furthermore, we have
\be\label{p-mul}
P_{ij} P_{kl} &=& \delta_{jk} P_{il} \nn
\sum_{i=1}^4 P_{ii} &=& \openone \nn
P\til_{11} &=& P_{11},
\ee
if $\til$ includes complex conjugation of the non-geometric
factor $i$. One may notice, that the conjugation used in
the Dirac adjoint is $\bar{~}$ and not $\til$. The $i$ and $j$
are not an index, but a label of the base elements. We can
construct a spinor base to represent the elements of
$CL(\openR,\eta)$. Therefore we set 
\be
\xi_i &:=& P_{i1}=\bar{t}_i P_{11} \nn
\xi^{\ast T}_i &:=& P_{1i} = P_{11} t_i.
\ee
Now, it is easily seen, that we have {\it four} possible choices
to construct such spinor representations, due to the four
idempotents $P_{ii}$, which yield {\it identical} matrix
representations. The above choice of the $t_i$ and $\bar{t}_i$
ensures that the spinor representations of the $e^\mu$ equals
the usual Dirac representation for vanishing $F^{\mu\nu}$. The
fourfold possibility is well known, see Parra \cite{Parra}.

Due to the multiplication rule (\ref{p-mul}), left
multiplication preserves this structure, while right
multiplication does mix the four possibilities. In this way, it
is not obvious, how one can connect ideal spinors and
Dirac--Hestenes spinors (elements of $CL^+$), which are build up
from all four spinor modules ($P_{ik}$, $k\in\{1,2,3,4\}$,
fixed) and are subjected to a right action, which indeed mixes
these modules. The explanation of iso-spin from the right action
of certain elements in the Dirac--Hestenes theory does thus
involve up to four {\it distinct}\/ spinor modules (particles).

The above developed method to start from the Grassmann algebra
and the $j$--sources give a {\it unique} construction to spinor
modules. The ambiguity is thereby removed, which is of extreme
importance in QFT, where the index-sets become infinite. Thus
one has not to bother, which of the four (infinite many in QFT)
modules to choose and how to establish their connection.

To establish this in Dirac theory, we have to translate equation
(\ref{Dirac}) into our picture. One would expect from
dimensional arguments, that the algebra $\Lambda[\openR^3]$
should be large enough to carry the Dirac spinor (each of them
has 8 real parameters). But, the mass term, with its
unconvenient feature to mix even and odd parts of the state
forces us to use $\Lambda[\openR^4]$, with four sources $\{
j_1,j_2,j_3,j_4 \}$. A ``state'' $X$ is the written as
($\alpha_{i\ldots}\in \openC$, s.c. employed) 
\be
X &=& \alpha_0\Id+\alpha_i j_i+\alpha_{ij,i<j}j_i\wedge j_j 
+\alpha_{ijk,i<j<k}j_i\wedge j_j\wedge j_k
+\alpha_{ijkl,i<j<k<l}j_i\ldots j_l.
\ee
The $\gamma$--matrices become via the definition
(\ref{def-gamma} or \ref{def-neu-gamma}) functions of
$\partial,j$. Furthermore, we have to search for a ``vacuum
state'', which has to fulfill the 
relation 
\be
\partial^x \vert 0>_F =0 &~~~&\forall x\in \V(1,3).
\ee
We could achieve this by 
\be
\vert 0>_F &=& \partial^1\partial^2\partial^3\partial^4,
\ee
but we prefer another choice, because we would like to require
$\vert 0>_F$ to be a (primitive) idempotent element, which
provides us an ``scalar product''. One should thus write $\vert
0>_F<0\vert$ to emphasize this feature. Hence we define
\be\label{fock}
\vert 0>_F<0\vert (\equiv \vert 0>_F \mbox{~loosely~})&:=&
\partial^1\partial^2\partial^3\partial^4 j_4 j_3 j_2 j_1.
\ee
This state obtains the properties
\be
(\vert 0>_F<0\vert )^2 &=& \vert 0>_F<0\vert \nn
X\vert 0>_F<0\vert &=& \vert X>_F<0\vert \nn
\vert 0>_F<0\vert X \vert 0>_F<0\vert &=& <X> \vert 0>_F<0\vert.
\ee
The functional picture, which is adapted to field theoretic
considerations, has a very close connection to Crumeyrolle's
construction of spinors \cite{Crumeyrolle90}. The distinction
arises from the arbitrary bilinear form $B$ and from the fact,
that the isotropic space was introduced only for technical
reasons. The development here is not restricted  to this special
case.

We can now define a base, which is large enough to carry a
representation of the $e^\mu$. This base is given as
\be
\vert \alpha > &\in& \{ j_\alpha \vert 0>_F<0\vert \},
\ee
where $\alpha$ is an ordered possible empty index set out of
$\{1,2,3,4\}$ and $j_\emptyset\equiv \Id$. If we represent the
base elements on them self, we obtain a ``spinor'' representation
\be
{}[\vert 0>_F<0\vert ] &=&
\left[ \begin{array}{ccc}%
1      & 0      & \cdots \\%
0      & \ddots & ~\\%
\vdots &    ~   &%
\end{array} \right]_{16\times 16} \nn
{}[j_1\vert 0>_F<0\vert ] &=&
\left[ \begin{array}{ccc}%
0      & 0      & \cdots \\%
1      & \ddots & ~\\%
\vdots &     ~  &%
\end{array} \right]_{16\times 16}
\ee
etc. The other matrices are given by terms like
\be
{}[\vert 0>_F<0\vert\partial^1 ] &=&
\left[ \begin{array}{ccc}%
0      & 1      & \cdots \\%
0      & \ddots & ~\\%
\vdots &      ~ &%
\end{array} \right]_{16\times 16},
\ee
which is a right action and thus {\it not} present in our
formalism. The above given set of the $P_{ij}$, can thus be
reobtained by identification of $i,j$ with $\alpha,\beta$ thus
\be
P_{ij} &\approx & j_\alpha \vert 0>_F<0\vert \partial^\beta.
\ee
The functional picture picks out one and only one of the spinor
representations. The $\beta$--index in (\ref{Dirac}) is thus of
the same kind as the above one. But due to the general
construction including non-trivial $F^{\mu\nu}$ we need the full
set of 16 elements and not 4 complex or 8 real ones. The
representation matrices of the functional bases show {\it no}\/
dependence of the metric $\eta^{\mu\nu}$ {\it nor}\/ of the
alternating part $F^{\mu\nu}$. This changes drastically if one
calculates matrix representations of $e^\mu$'s, which are highly
asymmetric and $\eta^{\mu\nu}$ as $F^{\mu\nu}$ dependent
\cite{Fauser-vertex}. $F^{\mu\nu}$ may be a function of space
and time, even if one works within a static Minkowsky space
($\eta^{\mu\nu}\equiv$ constant). 

To compare Dirac theory with QFT, it is convenient to use
Hamilton formalism. Formula (\ref{Dirac}) results in the
algebraic picture as ($\gamma^{0~-1}=\gamma^0$)
\be
i\frac{\partial}{\partial x^0} \psi &=&
\sum_{k=1}^3
(-\gamma^0\gamma^k \frac{\partial}{\partial x^k}
 -\frac{e}{\hbar c}\gamma^0\gamma^k A_k)\psi
-(\frac{e}{\hbar c} A_0 +mc\gamma^0)\psi.
\ee
The transition into the functional picture is obtained with
\be
\psi &\longrightarrow & X\vert 0>_F<0\vert \equiv \vert X> \nn
\gamma^\mu & \longrightarrow & \sum_\nu(B^{\mu\nu}j_\nu)
+\partial^\mu. 
\ee
Hence we may calculate the functional energy equation
\be
E\vert X> &=& i\frac{\partial}{\partial x^0} \vert X> =
H[j,\partial] \vert X>.
\ee
The functional Hamiltonian $H[j,\partial]$ requires long winded
calculations, which are not illuminating. But, comparing the
structural form of this Dirac functional equation to functional
equations of Dyson--Schwinger--Freese hierarchies in QFT
\cite{StumpfBorne} is quite interesting.
\begin{itemize}
\item 
The Hamilton formulation suggests to treat Dirac theory within
Pauli algebra ($CL_{3,0}$), which would reduce the dimension of
the state space to 8 \cite{Daviau}.
\item
The mass term, due to the $\gamma^0$ mixes odd and even parts of
the state. This is a very uncommon feature in QFT, where the
hierarchy equations decouple in odd and even ones. This is a
kind of ``super symmetry'', which is known to be relevant in
Dirac theory \cite{Thaller}. Therefore we identify the
Yvon--Takabayasi angle as a measure of particle number
non-conservation. 
\item
The functional sources and states provide a metric and
$F^{\mu\nu}$ independent set of base elements. In great
contrast we observe a $\eta^{\mu\nu}$ and $F^{\mu\nu}$ dependence
of the usual base elements $e^\mu$. Even in static Minkowsky
space $\openM_{1,3}$, there may be an up to now overlooked
space-time dependence due to a nontrivial
$F^{\mu\nu}=F(x^\mu,x^\nu)$. This plays an important role in
QFT, where $F^{\mu\nu}$ can be identified with the propagator
\cite{Fauser-pos,Fauser-thesis}.  
\item
To cure the unconvenient features of functional Dirac theory,
one should investigate from the beginning Dirac theory within
$CL_{4,1}$, where also the non-geometric $i$\/ is turned into a 
geometric entity. The functional state space becomes then 32
dimensional, which equals the real degrees of freedom in
complexified Dirac theory. Within this picture, there may be a
chance to reobtain full QED, which is based on {\it four}\/
Fock-like oscillator degrees of freedom, each described by
ordinary four-component spinor field operators
($a_\uparrow^\dagger, a_\downarrow^\dagger, d_\uparrow,
d_\downarrow$). This may also be the link to the fourfold
possibility obtained in Dirac--Hestenes theory by Parra. 
\item
In usual Dirac theory, one postulates {\it a priori}\/ the
connection between the spinor and its adjoint. Due to this, the
$F^{\mu\nu}$ is fixed (to zero). In our approach, the
$F^{\mu\nu}$ is not fixed and has to be calculated from the
theory. This requires a non-linear equation, which in QFT is
obtained by the coupling to the vacuum.
\item
Because it is possible to calculate the functional Hamiltonian
$H[j,\partial]$, one can ask for the ``one-particle'' theory,
which results in this functional ``field quantized'' equation.
This is the reversed question, how to quantize classical
(spinor) fields in QFT. One obtains such an equation by
``one-particle'' projections ${}_F<0\vert\partial^i$
\be
<0\vert \partial^i E\vert X> &=&
<0\vert \partial^i H[j,\partial] \vert X>\nn
E \alpha^i &=& H^i \alpha_0 + \sum H^{ij}\alpha_j +\ldots .
\ee
This equations may be non-linear.
\item
Since in Dirac theory one does require the scalar product to be
$\openC$--valued, we have for one state
\be
<\Psi \vert \Psi > = \bar{\Psi}^D\Psi 
&\rightarrow& \openR.
\ee
Thus $\Psi$ has to be in the Lipschitz group $\Gamma(p,q)$,
because only there we have
\be
\Phi_\varepsilon(\Psi,\Psi) &=&
\alpha_\varepsilon(\Psi)\Psi \rightarrow \openR,
\ee
and $\Psi$ is decomposable into one-vectors as selfadjoint to be
$\openR$- and not $\openC$-valued. This is not a usual
requirement in QFT and thereby new. This picture turns the state
$\vert \Psi>$ to be in a Grassmannian. The classification of
such manifolds is given in terms of Stiefel--Withney classes. An
approach to the Dirac theory in such a tetrad formalism (mobiles
of streamlines) is given by Kr\"uger \cite{Krueger-icte}.
\item
{}From the features of the Dirac theory one may be able to
construct adjoint (left) functional states, which then
constitutes a functional metric (on elements of the Lipschitz
group). This should be possible for non-linear theories (QCD,
NLJ-models) also and is thus beyond the current QFT development
and beyond current Fock space methods e.g. perturbation theory.
\end{itemize}

\section{Conclusion}

We developed a new method to study Dirac theory. This was
motivated to search for an analogous picture, which was
previously helpful in QFT. The Chevalley deformation provides a
tool, to fix not only the quadratic form $\Q$ of the Clifford
algebra in use, but also to fix the multi-vector structure. This
multi-vector structure is needed to build scalar products and
expectation values. The alternating (antisymmetric) part of the
chosen bilinear form is thus an important part of the theory,
even if it is usually absent. 

The Dirac theory was shown to behave in an unexpected way if
considered as a (toy) QFT. The mass term breaks the usual observed
splitting of QFT functional equations in even and odd parts and
breaks ``particle number'' conservation.
This term is thus the source of a well known \cite{Thaller}
super symmetry. The formulation of the theory suggests very
strongly, that one should study Dirac theory by means of the
Pauli algebra, see therefore Daviau \cite{Daviau}, which is of
course obscured by the mass term.

The possibility to have beside a constant Minkowsky metric a
position dependent alternating part, opens new ways to study
Dirac theory including vacuum effects. 

We have shown, that the requirement to use ``states'' in QM
involves new features, not observed in the operator formalism of
Dirac--Hestenes or QF theory. In this sense does the
Dirac--Hestenes theory bear {\it no} direct link to measurement
since one does not explicit calculate (or fix by physical
motivated assumptions) the $F^{\mu\nu}$
parameters, which therein are set implicitly zero.

\end{document}